%% file: main.tex
\documentclass[a4paper, oneside, twocolumn, notitlepage, 10pt]{extarticle_ecoc}

\usepackage{ecoc}
\usepackage[mode=tex]{standalone} 
\usepackage{subfigure}
\usepackage{dsfont}
\usepackage{float}
\usepackage{algorithm}
\usepackage[noend]{algpseudocode}
\usepackage{tikz,pgfplots}
\usepgfplotslibrary{fillbetween}
\usepgfplotslibrary{groupplots}
\usepackage{balance}
\usepackage{acro}
\input{acronym}

 \usepackage{mathtools}
 \usepackage{nccmath}

\usetikzlibrary{positioning}
\usetikzlibrary{calc}

\usepackage{booktabs,siunitx, multirow,ragged2e}

\newcommand{\vect}[1]{\ensuremath{\boldsymbol{#1}}}%
\newcommand*{\inlineequation}[2][]{%
  \begingroup
    \refstepcounter{equation}%
    \ifx\\#1\\%
    \else
      \label{#1}%
    \fi
    \relpenalty=10000 %
    \binoppenalty=10000 %
    \ensuremath{%
      #2%
    }%
    ~\@eqnnum
  \endgroup
}
\makeatother
\usepackage{hyperref}

\usepackage{paralist}
\usepackage{xcolor}
\tikzset{%
	partial ellipse/.style args={#1:#2:#3}{%
		insert path={+ (#1:#3) arc (#1:#2:#3)}%
	}%
}%

\begin{document}
\selectlanguage{american}    

\title{Blind Frequency-Domain Equalization Using\\ Vector-Quantized Variational Autoencoders}%

\vspace{-0.5cm}
\author{
Jinxiang Song\textsuperscript{(1)}, 
Vincent Lauinger\textsuperscript{(2)},
    Christian H\"{a}ger\textsuperscript{(1)},
    Jochen Schr\"{o}der\textsuperscript{(3)}, \\ Alexandre Graell i Amat\textsuperscript{(1)},
     Laurent Schmalen\textsuperscript{(2)}, and Henk Wymeersch\textsuperscript{(1)} 
}
\maketitle

\setstretch{1.1}
\renewcommand\footnotemark{}
\renewcommand\footnoterule{}

\vspace{2cm}
\begin{strip}
 \begin{author_descr}
 
   \textsuperscript{(1)}
   Dept.~of Electrical Engineering, Chalmers Univ.~of Technology, Sweden, \textcolor{blue}{\uline{jinxiang@chalmers.se}}$\!\!\!$\\
   \textsuperscript{(2)}
   Communications Engineering Lab, Karlsruhe Inst. of Technology, Germany \\
   \textsuperscript{(3)}
   Dept.~of Microtechnology and Nanoscience, Chalmers Univ.~of Technology, Sweden\\
   
 \end{author_descr}
\end{strip}

\begin{strip}
  \begin{ecoc_abstract}
    We propose a novel frequency-domain blind equalization scheme for coherent optical communications. The method is shown to achieve similar performance to its recently proposed time-domain counterpart with lower computational complexity, while outperforming  the commonly used CMA-based equalizers.  \textcopyright 2023 The Author(s)
  \end{ecoc_abstract}
\end{strip}

\begin{figure*}
    \centering
    \vspace{-1cm}
     \includegraphics[width=0.88\textwidth]{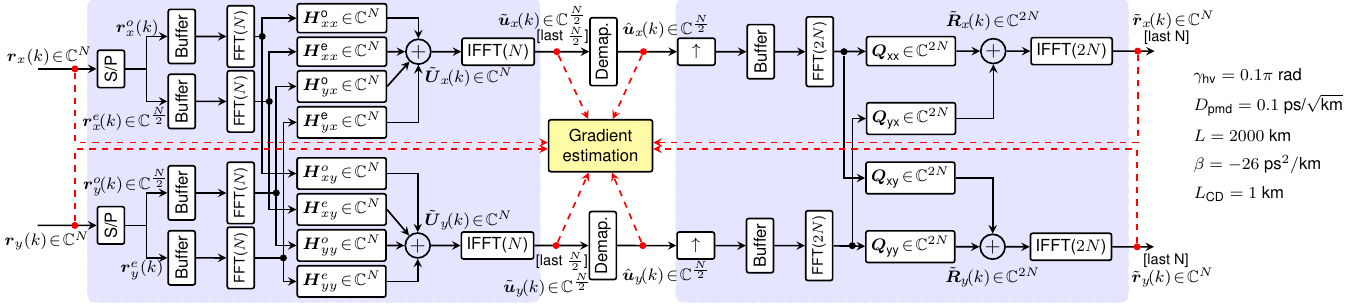}
    \caption{Block diagram of the proposed blind FD equalizer using the VQ-VAE framework. S/P: serial to parallel; Buffer: samples from the previous block are saved in the buffer and concatenated to the current block prior to the FFTs.}
    \label{fig:equalizer_model}
\end{figure*}

\section{Introduction}

In coherent optical communication systems with high-order modulation formats, data-aided \ac{DSP} algorithms
are commonly used for channel equalization\cite{spinnler2013pilot}, frequency offset estimation\cite{morsy2011feedforward}, and carrier phase recovery (CPR)\cite{6151032}, due to their
 modulation format independence, fast convergence, and high reliability. 

However, data-aided approaches decrease the overall system throughput as they require pilot symbols. Specifically, for channel equalization, which is the most pilot-demanding DSP block, the required overhead increases with the  channel memory length and desired tracking speed.
To circumvent this problem,
high-performance blind alternatives are preferable. The most popular algorithm for blind equalization is the \ac{CMA}\cite{godard1980self} and its variants\cite{525219}. Designed for constellations with constant signal amplitude, CMA generally converges  slowly for high-order modulation formats and is therefore typically combined with decision-directed (DD) algorithms \cite{fatadin2009a}.  DD algorithms, however, fail at low \ac{SNR} where decision errors lead to equalizer misconvergence.
As an alternative, a blind method  based on \acp{VAE} was recently proposed\cite{caciularu2020unsupervised}, showing superior performance than the \ac{CMA} for
high-order modulation formats\cite{lauinger2022blind}. 
Training of the VAE equalizer relies on maximizing a closed-form solution of the evidence lower bound (ELBO), which may be difficult to compute or might not even exist for some channels\cite{song2023blind}. To overcome this problem, we recently proposed the \ac{VQ-VAE} blind equalizer, showing improved performance over the VAE approach for both linear and nonlinear channels\cite{song2023blind}.


In this paper, we extend\cite{song2023blind} and present a novel \ac{FD} \ac{VQ-VAE} equalizer, with the aim of reducing computational complexity compared to its \ac{TD} alternative. 
Using \ac{FFT} and \ac{IFFT} filters~\cite{faruk2011adaptive},  FD-equalization closely matches the required block-based training of \ac{VAE}-based equalizers. In contrast to VAE equalizers, which have no known implementation in the FD, due to the reliance on the ELBO, 
the proposed 
VQ-VAE equalizer can readily be implemented in the FD, and exhibits fast and stable training behavior. It can reduce computational complexity compared to its \ac{TD} counterpart, especially in channels with long memories.
The performance of the proposed FD VQ-VAE equalizer is evaluated in a \ac{PDM} setup. Simulations show that our method achieves the same performance as the \ac{TD} equivalent while outperforming the widely used \ac{CMA} equalizer.

\section{Simulation Environment}
 We consider a single \ac{PDM} signal in a linear fiber channel with \ac{PMD}, residual chromatic dispersion (CD), and amplified spontaneous emission noise\cite{ip2007digital}. We consider block-based processing and a time-invariant channel  within one transmission block. The channel is modeled by the linear \ac{FD} channel matrix
\begin{align*}
     \boldsymbol{H}(f)=\boldsymbol{R}^\top  \begin{pmatrix}
      e^{j\pi\tau  f} & 0\\
     0 & e^{-j\pi\tau f}
     \end{pmatrix} \boldsymbol{R} \, e^{- j 2\pi^2 \beta L_\text{CD} f^2}, 
 \end{align*}
where 
$\smash{\boldsymbol{R}=\begin{psmallmatrix}\!\cos(\gamma) & \sin(\gamma)\\
-\sin(\gamma) & \cos(\gamma)\!\end{psmallmatrix}}$
represents a static rotation of the reference polarization to the fiber's \ac{PSP}, characterized by the parameter $\gamma$. The differential group delay between the two \acp{PSP} is denoted by 
$\smash{\tau= D_{\text{pmd}}\sqrt{L}}$, 
where $D_{\text{pmd}}$ is the \ac{PMD} parameter and $L$ is the fiber length. CD is assumed to be coarsely compensated before channel equalization and the residual dispersion is characterized by the equivalent  uncompensated fiber length denoted by $L_\text{CD}$.  

\section{FD Equalizers Based on VQ-VAEs}
A block diagram of the proposed blind \ac{FD} equalizer for \ac{PDM} transmission is depicted in Fig.~\ref{fig:equalizer_model}. Similar to\cite{song2023blind}, the proposed method  employs an encoder-decoder pair. The \emph{decoder} equalizes a block of $2\times$ oversampled channel observations $\smash{\boldsymbol{r}(k)=[\boldsymbol{r}_x(k), \boldsymbol{r}_y(k)] \in \mathbb{C}^{N\times 2}}$, where $N$ is the block size,
after which a demapper is employed to generate estimates $\smash{\hat{\boldsymbol{u}}(k)=[\hat{\boldsymbol{u}}_x(k), \hat{\boldsymbol{u}}_y(k)] \in \mathbb{C}^{\frac{N}{2}\times 2} }$ of the transmitted symbols $\smash{\boldsymbol{u}(k) =[\boldsymbol{u}_x(k), \boldsymbol{u}_y(k)]}$. The \emph{encoder}  estimates the channel and generates  reconstructions $\smash{\tilde{\boldsymbol{r}}(k) = [\tilde{\boldsymbol{r}}_x (k), \tilde{\boldsymbol{r}}_y(k)] \in \mathbb{C}^{N\times 2} }$  of the channel observations $\smash{\boldsymbol{r}(k)}$ from the estimated symbol vector $\smash{\hat{\boldsymbol{u}}(k)}$. 
In the following, we describe the \ac{FD} implementation of the {VQ-VAE} equalizer.

\noindent \emph{VQ-VAE Decoder:}
Fig.~\ref{fig:equalizer_model} (left) depicts the equalizer (decoder) model. 
The equalization is performed in the FD\cite{faruk2011adaptive}, using eight \ac{FD} filters. The eight FD filters are assigned into two groups, where each group forms a $\smash{2\times2}$ butterfly equalizer, denoted as the even and odd sub-equalizer, respectively.\footnote{ \ac{TD} equalizers typically operate at 2 samples per symbol to avoid frequency aliasing. Using sub-equalizers in FD naturally guarantees the equalized signal to have the correct rate.} In each time step $k$, a block of $N$ channel observations $\boldsymbol{r}(k)$ is first divided into even $\smash{\boldsymbol{r}^e(k)=[\boldsymbol{r}^e_x(k), \boldsymbol{r}^e_y(k)]\in \mathbb{C}^{\frac{N}{2} \times 2}}$ and odd observations $\smash{\boldsymbol{r}^o(k)=[\boldsymbol{r}^o_x(k), \boldsymbol{r}^o_y(k)]\in \mathbb{C}^{\frac{N}{2} \times 2}}$.
To enable fast implementation of linear convolution in \ac{FD}, we employ the overlap-save method with a $50\%$ overlap\cite{shynk1992frequency}. 
The even and odd channel observations are first concatenated to the buffered samples (i.e., $\smash{\boldsymbol{r}^e(k\!-\!1)}$ and $\smash{\boldsymbol{r}^o(k\!-\!1)}$) from the previous block and then converted into the \ac{FD} using \acp{FFT}. Then, the \ac{FD} signal is multiplied by the \ac{FD} filters of length $N$, resulting in the \ac{FD} equalized signal ${\tilde{\boldsymbol{U}}(k) = [\tilde{\boldsymbol{U}}_x(k), \tilde{\boldsymbol{U}}_y(k)] \in \mathbb{C}^{N\times 2}}$. The desired \ac{TD} equalized signal $\tilde{\boldsymbol{u}}\in \mathbb{C}^{\frac{N}{2}\times2}$ is then obtained by performing \acp{IFFT} and keeping the last $N/2$ samples.
Finally, a decision block that takes into account the prior distribution of the transmitted symbol (i.e., the maximum a posterior demapper) is employed to generate estimates $\hat{\boldsymbol{u}}(k)$ of the transmitted symbols ${\boldsymbol{u}}$. The entire decoder is denoted by $f_{\boldsymbol{\phi}}(\cdot)$, where $\boldsymbol{\phi}$ contains all trainable parameters (i.e., the parameters of the eight \ac{TD} filters) in the decoder block.\footnote{Note that FD equalizers are obtained from TD filters using FFTs so that TD and FD equalizers have the same degree of freedom. In practice, one can implement the equalizer in FD directly to fully exploit the advantages of FD equalization.} 

\noindent \emph{VQ-VAE Encoder:} As shown in Fig.~\ref{fig:equalizer_model} (right), the channel estimator (encoder) consists of 4 \ac{FD} filters that are connected in a $\smash{2\times2}$ butterfly structure. The estimated symbol vector $\hat{\boldsymbol{u}}$ is $2\times$ upsampled via zero insertion and concatenated to the buffered samples from the previous block before being converted into \ac{FD} using \ac{FFT} of size $2N$. Next, the FD filters are applied, after which \acp{IFFT} of size $2N$ are performed to get the TD reconstructed channel observations, finally keeping the last $N$ samples as $\tilde{\boldsymbol{r}}(k)$. We denote the entire encoder block by $f_{\boldsymbol{\theta}}(\cdot)$, where $\boldsymbol{\theta}$ denotes the set of all trainable parameters in the encoder block.

\begin{figure*}
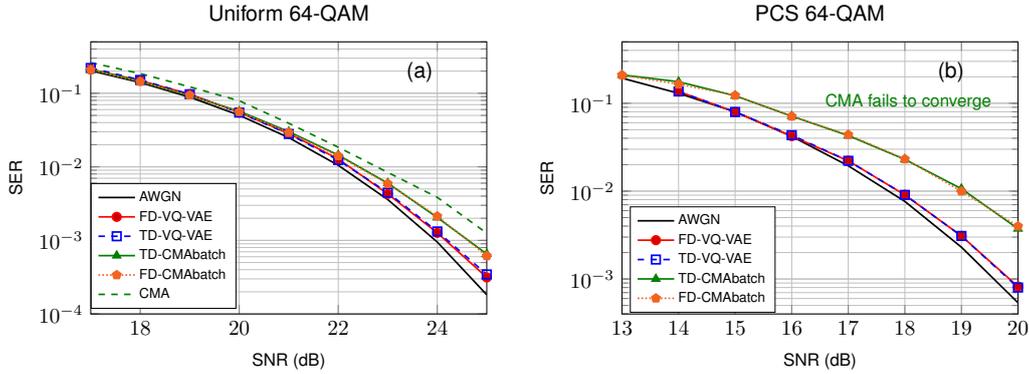

    \centering
    \vspace{-0.9cm}
     \includestandalone[width=0.85\textwidth]{figures/SER}
    \caption{SER for different SNRs: (a) transmission of unshaped $64$-QAM; (b) transmission of PCS-$64$-QAM ($\mathcal{H}=5$ bits/symbol)}
    \label{fig:SER}
    \vspace{-0.1cm}
\end{figure*}

\noindent \emph{VQ-VAE training:} Training  of the proposed equalizer is performed by minimizing
$\ell(\boldsymbol{\phi}, \boldsymbol{\theta}) = (1-\rho)\Vert \boldsymbol{r}-  \tilde{\boldsymbol{r}}  \Vert^2  + \rho \Vert \tilde{\boldsymbol{u}}-  \hat{\boldsymbol{u}}  \Vert^2$, which overcomes the challenges in standard VAE-based equalizer training (i.e.,  requiring a closed-form solution for the ELBO). The goal of the cost function is to push the decoder output to its nearest constellation point, while assuring a faithful reconstruction of the channel observation can be achieved using the encoder output. 
Here, $\tilde{\boldsymbol{r}} = f_{\boldsymbol{\theta}}(\hat{\boldsymbol{u}})$, $\hat{\boldsymbol{u}} = f_{\boldsymbol{\phi}}({\boldsymbol{r}})$, the dependence of $\boldsymbol{\phi}$ on the equalizer parameters is implicit through $\hat{\boldsymbol{u}}$, and $\smash{0 <\rho\leq 1}$ is a hyperparameter that can be tuned to improve the convergence behavior.

\section{Results}

\emph{Simulation Setup:} Dual-polarized $64$-QAM and $64$-QAM with \ac{PCS}  at symbol rate $\smash{R_s=90\,\text{GBaud}}$ is transmitted; the transmitted signal is $2\times$ oversampled and pulse-shaped with a root-raised cosine filter with $10\%$ roll-off; the fiber channel parameters are summarized on right of Fig.~\ref{fig:equalizer_model}. The PCS-$64$-QAM is generated according to the  Maxwell-Boltzmann distribution with a target entropy of $\smash{\mathcal{H}=5}$ bits/symbol.  For the equalizer configuration, we set all \ac{TD} equalizers to have $\smash{N_\text{tap}^{\text{TD}}=32}$ delay taps, whereas all the \ac{FD} sub-equalizers of length $N$ (same as the FFT size in the equalizer) are obtained from  \ac{TD} filters of $N_\text{tap}^{\text{FD}}=N_\text{tap}^{\text{TD}}/2=16$ by performing $N$-point FFT (so that the TD and FD equalizers have the same degree of freedom). 
The FFT/IFFT block size in the equalizer is set to $N=32$ for the unshaped signal and is increased to $N=64$ for the PCS constellation to improve the training stability. The FFT/IFFT block size in the  encoder is $2N$ as it works on the $2\times$ oversampled signal.
All equalizers are trained by minimizing their associated cost function using the Adam optimizer\cite{kingma2014adam}, where the learning rate is fixed to $0.0008$ (except for CMA, as described later). We use $\rho=0.5$ for VQ-VAE-based equalizer training.

\noindent \emph{SER versus SNR:} We start by evaluating the  performance of the proposed FD equalizer under different \ac{SNR} conditions. For comparison, we also provide the SER achieved by the TD VQ-VAE-based equalizer,  the CMA-based equalizer, and a variant of \ac{CMA} denoted as CMAbatch\cite{crivelli2013architecture, lauinger2022blind}; the batch size of CMAbatch is the same as the VQ-VAE.  To deal with the phase ambiguity,  a genie-aided CPR is employed for both CMA and CMAbatch. Furthermore, since CMA and CMAbatch are step-size sensitive\cite{yi2019joint}, a step-size scheduling scheme is employed to improve the convergence of the CMA and CMAbatch equalizer\cite{lauinger2022blind}.

The equalization performance is measured in terms of \ac{SER} and is visualized in Fig.~\ref{fig:SER} (a) for transmission of uniform $64$-QAM
and in Fig.~\ref{fig:SER} (b) for transmission of PCS-$64$-QAM.  For both uniform $64$-QAM and PCS-$64$-QAM, the proposed FD equalizer is shown to achieve the same performance as its \ac{TD} counterpart, while outperforming the CMA and CMAbatch, especially for \ac{PCS} constellation.

\begin{figure}
    \centering
     \includegraphics[width=0.9\columnwidth]{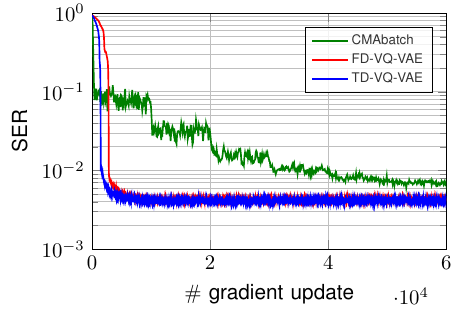}
    \caption{SER versus the number of gradient updates. The number of samples used in all algorithms is 32.}
    \label{fig:SER_evolution}
    \vspace{-0.2cm}
\end{figure}
\noindent\emph{Training Complexity:} 
The VQ-VAE-based equalizer requires additional computation compared to conventional equalizers,  to reconstruct the channel observation from the decoded symbols.  However, once converged, the VQ-VAE-based equalizer can be switched to the DD mode, i.e., by setting $\rho=1$. 
To capture the ability to 
adapt quickly to changing channel conditions, 
we measure the training complexity in terms of the number of gradient updates required for the equalizer to converge.
In Fig.~\ref{fig:SER_evolution}, we visualize the \ac{SER} versus the number of gradient updates for different algorithms when unshaped 64-QAM is transmitted. Results show that the proposed FD VQ-VAE-based equalizer has a similar convergence rate to its TD counterpart while converging much faster than CMAbatch, despite that step-size scheduling is employed for CMAbatch. 

\noindent\emph{Inference Complexity:} 
We quantify the inference complexity in terms of the number of complex-valued multiplications required for each equalized symbol per polarization. The  complexity is $2N_\text{tap}$ for the TD VQ-VAE equalizer and is $3\log_2N +8$ for the FD VQ-VAE equalizer, where $N$ is preferably set to $N_\text{tap}/2$. Hence, for $N_\text{tap}\ge 8$, the FD equalizer is preferable\cite{faruk2011adaptive}.

\section{Conclusion}
We have introduced a new approach to FD equalization using the VQ-VAE framework. Numerical results indicate that the proposed FD VQ-VAE-based equalizer retains all the desired properties of  its \ac{TD} counterpart  (fast convergence, ability to operate with PCS and higher-order modulation), while exhibiting lower complexity. 

\section{Acknowledgements}
\footnotesize{This work was supported by the Knut and Alice Wallenberg Foundation, grant No.~2018.0090, 
the Swedish Research Council under grant No.~2020-04718, 
and the German BMBF under grant No.~16KIS1316.}


\balance
\newpage 
\printbibliography

\end{document}

%% file: Acronym.tex
\DeclareAcronym{AE}{
short=AE,
long= autoencoder,
}

\DeclareAcronym{WDM}{
short=WDM,
long=wavelength division multiplexing ,
}

\DeclareAcronym{NN}{
short=NN,
long= neural network,
}
\DeclareAcronym{DSP}{
short=DSP,
long= digital signal processing,
}
\DeclareAcronym{CMA}{
short=CMA,
long= constant-modulus algorithm,
}

\DeclareAcronym{MMA}{
short=MMA,
long= multi-modulus algorithm,
}

\DeclareAcronym{FD}{
short=FD,
long= frequency-domain,
}
\DeclareAcronym{TD}{
short=TD,
long= time-domain,
}

\DeclareAcronym{PS}{
short= PS,
long= pulse-shaping,
}

\DeclareAcronym{DAC}{
short=DAC,
long= digital-to-analog converter,
}

\DeclareAcronym{ADC}{
short=ADC,
long= analog-to-digital converter,
}

\DeclareAcronym{DL}{
short=DL,
long= deep learning,
}

\DeclareAcronym{ENOB}{
short=ENOB,
long= effective number of bit,
}

\DeclareAcronym{DPD}{
short= DPD,
long= digital pre-distortion,
}

\DeclareAcronym{}{
short=,
long= ,
}

\DeclareAcronym{MZM}{
short= MZM,
long= Mach-Zehnder modulator ,
}

\DeclareAcronym{MZI}{
short= MZM,
long= Mach-Zehnder interferometer ,
}
\DeclareAcronym{ML}{
short= ML,
long= maximum-likelihood  ,
}
\DeclareAcronym{VAE}{
short= VAE,
long= variational autoencoder  ,
}

\DeclareAcronym{EDFA}{
short= EDFA,
long= erbium-doped fiber amplifier ,
}

\DeclareAcronym{SE}{
short= SE,
long= spectral efficiency ,
}

\DeclareAcronym{PA}{
short= PA,
long= power amplifier ,
}

\DeclareAcronym{MF}{
short= MF,
long= matched filter ,
} 

\DeclareAcronym{OOB}{
short= OOB,
long= out-of-band ,
}

\DeclareAcronym{ICI}{
short= ICI,
long= inter-channel interference,
}

\DeclareAcronym{IQM}{
short= IQM,
long= in-phase and quadrature modulator,
}

\DeclareAcronym{RRC}{
short= RRC,
long= root-raised-cosine,
}

\DeclareAcronym{AWGN}{
short= AWGN,
long= additive white Gaussian noise,
}

\DeclareAcronym{SER}{
short= SER,
long= symbol error rate,
}

\DeclareAcronym{RL}{
short= RL,
long= reinforcement learning,
}

\DeclareAcronym{SNDR}{
short= SNDR,
long= signal-to-noise-plus-distortion ratio,
}

\DeclareAcronym{SNR}{
short= SNR,
long= signal-to-noise ratio,
}

\DeclareAcronym{FIR}{
short= FIR,
long= finite impulse response,
}

\DeclareAcronym{MAP}{
short= MAP,
long= maximum a posterior 
}

\DeclareAcronym{PMD}{
short= PMD,
long= polarization mode dispersion 
}

\DeclareAcronym{PSP}{
short= PSP,
long=   principle state of polarization
}

\DeclareAcronym{FFT}{
short= FFT,
long=  fast Fourier transform
}

\DeclareAcronym{DFT}{
short= DFT,
long=  discrete Fourier transform
}
\DeclareAcronym{IFFT}{
short= IFFT,
long= inverse  FFT
}

\DeclareAcronym{IDFT}{
short= IDFT,
long= inverse  DFT
}

\DeclareAcronym{PCS}{
short= PCS,
long= probabilistic constellation shaping
}

\DeclareAcronym{VQ-VAE}{
short= VQ-VAE,
long= vector-quantized VAE
}

\DeclareAcronym{PDM}{
short= PDM,
long= polarization-division multiplexing
}

%% file: figures/SER.tex
\begin{tikzpicture}
\begin{semilogyaxis}[%
    scale only axis,
	xlabel= SNR (dB) , ylabel= SER,
	title = Uniform 64-QAM,
	xmin=17, xmax=25,%
	ymin=1.e-4, ymax=5e-1,%
	grid=both,%
 	width=6.5cm, height=4.5cm,%
	legend cell align={left},%
	label style={font=\footnotesize},
	tick style={font=\footnotesize},
	legend style={at={(0.0,0.25)},anchor=west,nodes={scale=0.8, transform shape}, font=\footnotesize,  fill=white}%
]%

\addplot +[thick, color=black, solid, mark=None] 
table {figures/AWGN_SER.txt};
\addlegendentry{AWGN}

\addplot +[thick, color=red, solid, mark=*] 
table {figures/FD_VQ_VAE_SER_32.txt};
\addlegendentry{FD-VQ-VAE}

\addplot +[thick, color=blue, dashed, mark=square, mark options={scale=1,solid }] 
table {figures/TD_VQ_VAE_SER_32.txt};
\addlegendentry{TD-VQ-VAE}

\addplot +[thick, color=green!50!black, solid, mark=triangle*] 
table {figures/TD_CMAbatch_SER_32.txt};
\addlegendentry{TD-CMAbatch}

\addplot +[thick,  color=yellow!25!red, densely dotted, mark=pentagon*,  mark options={scale=1,solid }] 
table {figures/FD_CMAbatch_32.txt};
\addlegendentry{FD-CMAbatch}

\addplot +[thick, color=green!50!black, dashed , mark=None] 
table {figures/CMA_SER.txt};
\addlegendentry{CMA}

\end{semilogyaxis}
\node at (5.4, 4.0) {(a)};
\end{tikzpicture}

\hspace{5mm}

\begin{tikzpicture}
\begin{semilogyaxis}[%
    scale only axis,
	xlabel= SNR (dB) , ylabel= SER,
	title = PCS 64-QAM ,
	xmin=13, xmax=20,%
	ymin=4.e-4, ymax=5e-1,%
	grid=both,%
 	width=6.5cm, height=4.5cm,%
	legend cell align={left},%
	label style={font=\footnotesize},
	tick style={font=\footnotesize},
	legend style={at={(0.02,0.2)},anchor=west,nodes={scale=0.8, transform shape}, font=\footnotesize,  fill=white}%
]%

\addplot +[thick, color=black, solid, mark=None] 
table {figures/PCS_H5.txt};
\addlegendentry{AWGN}

\addplot +[thick, color=red, solid, mark=*] 
table {figures/FD-VQ-VAE-SER_64_PCS.txt};
\addlegendentry{FD-VQ-VAE}

\addplot +[thick, color=blue, dashed, mark=square, mark options={scale=1,solid }] 
table {figures/TD_VQ_VAE_SER_64_PCS.txt};
\addlegendentry{TD-VQ-VAE}

\addplot +[thick, color=green!50!black, solid, mark=triangle*] 
table {figures/TD_CMAbatch_SER_32_PCS_H5.txt};
\addlegendentry{TD-CMAbatch}

\addplot +[thick, color=yellow!25!red, densely dotted, mark=pentagon*,  mark options={scale=1,solid }] 
table {figures/FD_CMAbatch_64_PCS_H5.txt};
\addlegendentry{FD-CMAbatch}

\end{semilogyaxis}
\node at (5.4, 4.0) {(b)};

\node[] at (4.7, 3.5) {\textcolor{green!50!black}{ \footnotesize{CMA fails to converge}}};
\end{tikzpicture}

%% file: references.bib
@article{lauinger2022blind,
  title={Blind equalization and channel estimation in coherent optical communications using variational autoencoders},
  author={Lauinger, Vincent and Buchali, Fred and Schmalen, Laurent},
  journal={IEEE J. Sel. Areas Commun.},
  volume={40},
  number={9},
  pages={2529--2539},
  year={2022}
}

@article{fatadin2009a,
  title = {Blind {{Equalization}} and {{Carrier Phase Recovery}} in a 16-{{QAM Optical Coherent System}}},
  author = {Fatadin, Irshaad and Ives, David and Savory, Seb J.},
  year = {2009},
  month = aug,
  journal = {J. Lightw. Techn.},
  volume = {27},
  number = {15},
  pages = {3042--3049},
}

@article{song2023blind,
  title={Blind Channel Equalization Using Vector-Quantized Variational Autoencoders},
  author={Song, Jinxiang and Lauinger, Vincent and Wu, Yibo and H{\"a}ger, Christian and Schr{\"o}der, Jochen and Schmalen, Laurent and Wymeersch, Henk and others},
  journal={arXiv preprint arXiv:2302.11687},
  year={2023}
}

@article{caciularu2020unsupervised,
  title={Unsupervised linear and nonlinear channel equalization and decoding using variational autoencoders},
  author={Caciularu, Avi and Burshtein, David},
  journal={IEEE Trans.  Cogn. Commun.  Netw.},
  volume={6},
  number={3},
  pages={1003--1018},
  year={2020},
  publisher={IEEE}
}

@article{morsy2011feedforward,
  title={Feedforward carrier recovery via pilot-aided transmission for single-carrier systems with arbitrary {M-QAM} constellations},
  author={Morsy-Osman, Mohamed and Zhuge, Qunbi and Chen, Lawrence R and Plant, David V},
  journal={Opt. Express},
  volume={19},
  number={24},
  pages={24331--24343},
  year={2011},
  publisher={Optica Publishing Group}
}

@ARTICLE{6151032,
  author={Magarini, Maurizio and Barletta, Luca and Spalvieri, Arnaldo and Vacondio, Francesco and Pfau, Timo and Pepe, Marianna and Bertolini, Marco and Gavioli, Giancarlo},
  journal={IEEE Photon. Technol. Lett.}, 
  title={Pilot-Symbols-Aided Carrier-Phase Recovery for {100-G PM-QPSK} Digital Coherent Receivers}, 
  year={2012},
  volume={24},
  number={9},
  pages={739-741}}

@INPROCEEDINGS{525219,
  author={Kil Nam Oh and Yong Ohk Chin},
  booktitle={Proc. ICC}, 
  title={Modified constant modulus algorithm: blind equalization and carrier phase recovery algorithm}, 
  year={1995},
  volume={1},
  number={},
  pages={498-502 vol.1}
  }

@article{godard1980self,
  title={Self-recovering equalization and carrier tracking in two-dimensional data communication systems},
  author={Godard, Dominique},
  journal={IEEE Trans. Commun.},
  volume={28},
  number={11},
  pages={1867--1875},
  year={1980}
}

@article{yi2019joint,
  title={Joint equalization scheme of ultra-fast {RSOP} and large {PMD} compensation in presence of residual chromatic dispersion},
  author={Yi, Wei and Zheng, Zibo and Cui, Nan and Zhang, Xiaoguang and Qiu, Liyuan and Zhang, Nannan and Xi, Lixia and Zhang, Wenbo and Tang, Xianfeng},
  journal={Opt. Express},
  volume={27},
  number={15},
  pages={21896--21913},
  year={2019},

}

@inproceedings{spinnler2013pilot,
  title={Pilot-assisted channel estimation methods for coherent receivers},
  author={Spinnler, Bernhard and Calabr{\`o}, Stefano and Kuschnerov, Maxim},
  booktitle={Proc. OFC},
  pages={1--3},
  year={2013}
}

@article{faruk2011adaptive,
  title={Adaptive frequency-domain equalization in digital coherent optical receivers},
  author={Faruk, Md Saifuddin and Kikuchi, Kazuro},
  journal={Opt. Express},
  volume={19},
  number={13},
  pages={12789--12798},
  year={2011},
}

@article{crivelli2013architecture,
  title={Architecture of a single-chip 50 {G}b/s {DP}-{QPSK}/{BPSK} transceiver with electronic dispersion compensation for coherent optical channels},
  author={Crivelli, Diego E and Hueda, Mario R and Carrer, Hugo S and Del Barco, Mart{\'\i}n and L{\'o}pez, Ramiro R and Gianni, Pablo and Finochietto, Jorge and Swenson, Norman and Voois, Paul and Agazzi, Oscar E},
  journal={IEEE Trans. Circuits Syst. I: Reg. Papers},
  volume={61},
  number={4},
  pages={1012--1025},
  year={2013}
}

@article{kingma2014adam,
  title={Adam: A method for stochastic optimization},
  author={Kingma, Diederik P and Ba, Jimmy},
  journal={arXiv preprint arXiv:1412.6980},
  year={2014}
}

@article{shynk1992frequency,
  title={Frequency-domain and multirate adaptive filtering},
  author={Shynk, John J and others},
  journal={IEEE Signal Process. Mag.},
  volume={9},
  number={1},
  pages={14--37},
  year={1992}
}

@article{ip2007digital,
  title={Digital equalization of chromatic dispersion and polarization mode dispersion},
  author={Ip, Ezra and Kahn, Joseph M},
  journal={J. Light. Techn.},
  volume={25},
  number={8},
  pages={2033--2043},
  year={2007},
}
